\documentclass{osa-article}

\journal{osajournal}

\articletype{Research Article}

\usepackage{graphicx}
\usepackage{bm}
\usepackage{hyperref}
\usepackage[english]{babel}

\usepackage{siunitx}
\usepackage{bbm}
\usepackage{cleveref}
\usepackage{esdiff}
\usepackage{tikz}

\DeclareMathAlphabet{\mathsfit}{\encodingdefault}{\sfdefault}{m}{sl}
\SetMathAlphabet{\mathsfit}{bold}{\encodingdefault}{\sfdefault}{bx}{sl}
\newcommand{\I}{\ensuremath{\mathrm{i}}}

\newcommand{\e}{\ensuremath{\mathrm{e}}}
\newcommand{\ee}{\ensuremath{\bm{\hat{e}}}}

\newcommand{\ez}{\ensuremath{\bm{\hat{z}}}}

\newcommand{\T}{\ensuremath{\bm{\mathsfit{T}}}}
\newcommand{\Cmat}{\ensuremath{\bm{\mathsfit{C}}}}
\newcommand{\Q}{\ensuremath{\bm{\mathsfit{Q}}}}

\newcommand{\PW}{PW}
\newcommand{\VSW}{VSW}
\newcommand{\TIO}{$\text{TiO}_\text{2}$}
\newcommand{\xeon}{Intel Xeon X5570}

\definecolor{dorange}{HTML}{FF7f0E}
\definecolor{dgreen}{HTML}{2CA02C}

\crefname{figure}{Fig.}{Figures}
\crefname{equation}{Eq.}{equations}

\begin{document}

\title{Efficient simulation of bi-periodic, layered structures based on the T-matrix method}

\author{Dominik Beutel,\authormark{1,*} Achim Groner,\authormark{1}, Carsten Rockstuhl,\authormark{1,2} and Ivan Fernandez-Corbaton\authormark{2}}

\address{\authormark{1}Institut f\"{u}r Theoretische Festk\"{o}rperphysik, Karlsruhe Institute of Technology, 76131 Karlsruhe, Germany\\
\authormark{2}Institute of Nanotechnology, Karlsruhe Institute of Technology, 76021 Karlsruhe, Germany}

\email{\authormark{*}Corresponding author: dominik.beutel@kit.edu} 



\begin{abstract}
Predicting the optical response of macroscopic arrangements of individual scatterers is a computational challenge, as the problem involves length scales across multiple orders of magnitude. We present a full-wave optical method to highly efficiently compute the scattering of light at objects that are arranged in bi-periodic arrays. Multiple arrays or homogeneous thin-films can be stacked to build up an entire multicomposite material in the third dimension. The scattering properties of the individual objects in each array are described by the T-matrix formalism. Therefore, arbitrarily shaped objects and even molecules can be the basic constituent of the arrays. Taking the T-matrix of the individual scatterer as the point of departure allows to explain the optical properties of the bulk material from the scattering properties of its constituents. We use solutions of Maxwell's equations with well defined helicity. Therefore, chiral media are particularly easy to consider as materials for both scatterers or embedding media. We exemplify the efficiency of the algorithm with an exhaustive parametric study of anti-reflective coatings for solar cells made from cylinders with a high degree of helicity preservation. The example shows a speed-up of about 500 with respect to finite-element computations. A second example specifically exploits the use helicity modes to investigate the enhancement of the circular dichroism signal in a chiral material.
\end{abstract}


\section{Introduction}\label{sec:intro}
The design and fabrication of artificial photonic materials with predefined functionalities is a major ongoing scientific and technological endeavor \cite{kadic2019,kuznetsov2016}. Advances in fabrication techniques currently allow the precise fashioning of bi-periodic arrays of inclusions, often called metasurfaces, or regular scaffolds of (chiral) molecules of macroscopic sizes \cite{lalanne2017,sain2019,heinke2019}. Fully three-dimensional artificial photonic materials with tailor-made properties \cite{barner-kowollik2017,gissibl2016} could soon be available thanks to 3D laser nanoprinting technologies. The accurate and efficient computation of the properties of artificial materials is crucial for the research and development of such complex systems. 

The electromagnetic response of artificial photonic materials is dictated by Maxwell equations which lack an analytical solution in most cases. We must then resort to numerical computations. Moreover, the fine details of artificial photonic materials with critical length scales comparable to one wavelength deny any simplifying approximations. For example, the typical electromagnetic size of ``metamolecules'' that compose 2D metasurfaces and 3D metamaterials prevent us from using the dipolar approximation. The individual scattering response is too complex, and much of the functionality actually emerges from this complexity. This complexity in the optical response of an individual scatterer also denies the homogenization of such materials with local constitutive relations, which, when applicable, simplify the further treatment \cite{smith2005}. The possibility of describing photonic metamaterials by means of non-local constitutive relations is currently under study \cite{mnasri2018,mnasri2019}. 

General purpose Maxwell solvers can faithfully simulate any system as long as each of its individual elements, e.g. individual ``metamolecules'', can be accurately described using the effective field approximation implicit in the material constitutive relations. Several such solvers implementing finite-element or finite-difference schemes are available \cite{burger2008,comsol,oskooi2010}. Their main drawback is a large computational cost. This limits the range of parametric studies, and their usability for optimization techniques that rely on a repeated response evaluation \cite{popov2014}. Especially, an efficient simulation method is crucial in the emerging field of inverse design for optical applications\cite{molesky2018,zhan2018,zhan2019}. Semi-analytical methods based on the Green Tensor are inherently more efficient and are being developed in the context of two-dimensional arrays of inclusions \cite{lunnemann2013,babicheva2017,swiecicki2017,swiecicki2017b,babicheva2018,alaee2017}. They can be seen as successive extensions of the electric dipole approximation by including the magnetic dipole and the electric and magnetic quadrupole terms. Such piecewise extensions have some limitations. For example, the inclusion of other layers made from scattering particles or even just the consideration of a homogeneous slab, like a substrate or a superstrate, is not straightforward. Also, the analytical complexity of adding each new term increases rapidly, which, for example, can complicate the treatment of recent developments geared to achieve scatterers with a well defined and resonant response in higher order multipoles \cite{zenin2020,terekhov2019}.  

These problems can be solved by using the T-matrix based \cite{mishchenko2017} computation of array responses \cite{stefanou1998,stefanou2000,gantzounis2006}. Essentially, this technique also relies on the Green Tensor, but the use of the T-matrix greatly facilitates the inclusion of arbitrarily-high multipolar terms, limited only by the available numerical capabilities. Additionally, the work from \cite{stefanou1998,stefanou2000,gantzounis2006} shows how the framework allows to model systems with an arbitrary number of layers. Nevertheless, the publicly available codes in \cite{stefanou1998,stefanou2000,gantzounis2006} are limited to arrays composed of spherical inclusions and non-chiral materials for the additional layers. For arrangements of a finite number of particles, the T-matrix method has already been proven to allow for efficient simulations \cite{egel2017,egel2017a}.

In here, we present an algorithm to rigorously compute the electromagnetic response from structures that are periodic in two dimensions and layered in the remaining third dimension. The building blocks of the periodic structures are described by T-matrices \cite{waterman1965}, which allows the consideration of particles of any shape, and even molecules and molecular ensembles \cite{fernandez-corbaton2020}. In addition to structures periodic in two dimensions, thick bulks of three-dimensional periodic structures can be simulated efficiently, bridging several orders of magnitude from the size of the single building block to the bulk material. The only assumption that we need to formulate this algorithm is the 2D periodicity of the planar arrays of scatterers. Each of them must be reproducible by the repetition of a unit cell on a Bravais lattice. The unit cell may contain different kinds of individual particles. Along the third dimension of the layered structure, however, each layer might consists of different scatterers or it can consist of a slab made from a homogeneous medium. Still, all layers are required to share the same periodicity.  

Our formulation uses solutions of the Helmholtz wave equation with well-defined helicity as the basis set \cite{birula1996,fernandez-corbaton2012}. This enables the straightforward consideration of chiral media \cite{lakhtakia1994,bohren1974} for the objects described by the T-matrices, for the embedding media, and for any additional layers. We show the ability to handle chiral media in an example where spheres are used for enhancing the circular dichromism signal of a chiral medium. The analytic expressions for multi-scattering processes described by T-matrices \cite{mishchenko1996}, and the use of methods for rapid convergence of lattice sums \cite{kambe1967}, makes the algorithm two to three orders of magnitude faster than finite element methods. We show such speed-up factor in an exemplary application, where we perform an exhaustive parametric study of nano-structured anti-reflective coatings for solar cells. 

In the rest of the article, we first start by constructing plane wave (PW) solutions for the wave propagation in periodic chiral media. These solutions are used to describe the scattering from single periodic layers or at flat interfaces. The latter include interfaces between different chiral media. We calculate the scattering within periodic arrays of scatterers by a transition from the PW basis to a basis with vector spherical waves (VSWs) of well-defined helicity. The multi-scattering process is described by the translation coefficients for VSWs \cite{cruzan1962,stein1961}. We perform the necessary summation of such coefficients by using the Ewald summation method for two-dimensional media \cite{kambe1967}, which results in rapidly converging sums. Finally, another such sum gives the transition from scattered VSWs back to PWs \cite{popov2014}. We then provide two exemplary application of the algorithm. First, we study nano-structured anti-reflective coatings for solar cells. The coating consists of cylinders in a periodic array on top of a solar cell. We analyze the influence of the lattice geometry and the direction of incidence on the anti-reflective behavior by means of parameter sweeps: A total of almost 250.000 simulations that took less than two days on an Intel Xeon X5570 workstation with our code, which would not be practically feasible using a referential finite-element solver \cite{burger2008}. Second, we demonstrate the simulation of structures consisting of chiral materials by an example where spheres immersed in a chiral solution are used to  enhance the circular dichroism (CD) signal.

\begin{figure}%
	\centering
	\includegraphics{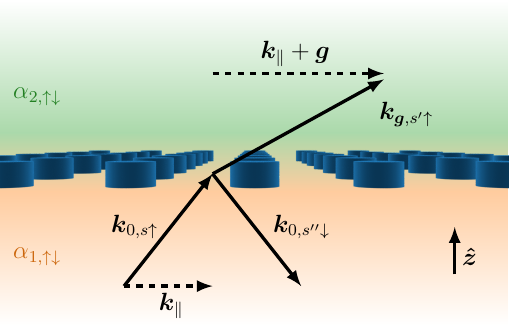}
	\caption{Sketch of the scattering of an incident plane wave with wave vector $\bm{k}_{0s\uparrow}$ at a two-dimensional periodic structure. The wave vector has the tangential component $\bm{k}_\parallel$. The array lies in the x-y-plane, which separates the space in two halves, where the field is expanded according to \cref{eq:pwexp,eq:Qgeneral} with either $\alpha_{1,\uparrow\downarrow}$ ($z < 0$) or $\alpha_{2,\uparrow\downarrow}$ ($z > 0$). In the figure, the wave vector $\bm{k}_{\bm{g}, s' \uparrow}$ of a specific diffraction order $\bm{g}$ is shown in transmittance and the zeroth order wave vector $\bm{k}_{0, s'' \downarrow}$ in reflectance.}\label{fig:sketch}
\end{figure}

\section{Calculation method}\label{sec:method}
We consider structures consisting of periodically arranged scatterers in two spatial dimensions and that are layered in the third dimension. The method is based on eigenmodes of the helicity operator and, thus, it suits chiral media \cite{lakhtakia1994}. The embedding medium of these scatterers can be stratified with the media interfaces located between different layers of scatterers. Therefore, this method is suitable for a wide range of objects, including three-dimensional metamaterials, photonic crystals, and metasurfaces on thin films and substrates. For the calculation, the whole structure is separated into individual arrays of scatterers and planar interfaces, for which the scattering problem is solved individually \cite{stefanou1998,stefanou2000,gantzounis2006}. In a subsequent step, these layers are assembled back into the whole structure, taking into account the multiple scattering between them \cite{stefanou1998,stefanou2000,gantzounis2006}.

Hereinafter, we will first introduce the method to describe the structure in a layer-by-layer manner with a set of four matrices, which we call Q-matrices  \cite{stefanou1998,stefanou2000}. Then, we will describe how to explicitly obtain the matrix entries for a bulk material and for an infinitely extended interface. Finally, we will solve the multiple-scattering problem in an infinitely extended array to obtain the Q-matrix entries for it.

\subsection{Layer description and coupling}
To begin with, we consider a single layer of scatterers, as shown in \cref{fig:sketch}, with a finite extend in the z-direction. The layer divides the domain of the scattering problem into two half-spaces below ($j=1$) and above ($j=2$) it. Each of the half-spaces is filled with an isotropic homogeneous material characterized by a relative permittivity $\epsilon_j(\omega)$, relative permeability $\mu_j(\omega)$, and chirality parameter $\kappa_j(\omega)$.
Using these parameters, the constitutive equations relating the monochromatic electric and magnetic fields $\bm{E}_j(\bm{r},\omega)$ and $\bm{H}_j(\bm{r},\omega)$ and flux densities $\bm{D}_j(\bm{r},\omega)$ and $\bm{B}_j(\bm{r},\omega)$ are taken to be \cite{kristensson2016}
\begin{equation}
	\begin{pmatrix}
		\frac{1}{\epsilon_0}\bm{D}(\bm{r},\omega) \\
		c_0\bm{B}(\bm{r},\omega)
	\end{pmatrix}
	=
	\begin{pmatrix}
		\epsilon(\omega) & \I\kappa(\omega) \\
		-\I\kappa(\omega) & \mu(\omega)
	\end{pmatrix}
	\begin{pmatrix}
		\bm{E}(\bm{r},\omega) \\
		Z_0\bm{H}(\bm{r},\omega)
	\end{pmatrix}\,,
\end{equation}
where we used the constants $c_0$ for the speed of light in vacuum, $\epsilon_0$ for the vacuum permittivity, and $Z_0$ for the vacuum impedance. Also, the $j$ subindex has been dropped to avoid clutter.
From these constitutive relations combined with the Maxwell equations, we obtain
\begin{equation}
	\begin{split}
		\bm{\nabla}\times
		\begin{pmatrix}
			\bm{E}(\bm{r},\omega) \\
			Z_0\bm{H}(\bm{r},\omega)
		\end{pmatrix}
		=\frac{\omega}{c_0}
		\begin{pmatrix}
			0 & \I \\
			-\I & 0
		\end{pmatrix}
		\begin{pmatrix}
			\frac{1}{\epsilon_0}\bm{D}(\bm{r},\omega) \\
			c_0\bm{B}(\bm{r},\omega)
		\end{pmatrix} \\
		=\frac{\omega}{c_0}
		\begin{pmatrix}
			\kappa(\omega) & \I\mu(\omega) \\
			-\I\epsilon(\omega) & \kappa(\omega)
		\end{pmatrix}
		\begin{pmatrix}
			\bm{E}(\bm{r},\omega) \\
			Z_0 \bm{H}(\bm{r},\omega)
		\end{pmatrix}
	\end{split}
\end{equation}
which can be diagonalized to
\begin{equation}
	\bm{\nabla}\times
	\begin{pmatrix}
		\bm{G}_+(\bm{r},\omega) \\
		\bm{G}_-(\bm{r},\omega)
	\end{pmatrix}
	=
	\begin{pmatrix}
		k_+(\omega) & 0 \\
		0 & -k_-(\omega)
	\end{pmatrix}
	\begin{pmatrix}
		\bm{G}_+(\bm{r},\omega) \\
		\bm{G}_-(\bm{r},\omega)
	\end{pmatrix}\label{eq:chiral}
\end{equation}
with $c_0k_\pm(\omega) = \omega\left(\sqrt{\epsilon(\omega)\mu(\omega)}\pm\kappa(\omega)\right)$ and the Riemann-Silberstein vectors \cite{silberstein1907,birula1996} $\sqrt{2}\bm{G}_\pm(\bm{r},\omega) = \bm{E}(\bm{r},\omega) \pm \I Z_0Z(\omega)\bm{H}(\bm{r},\omega)$. $Z(\omega) = \sqrt{\mu(\omega)/\epsilon(\omega)}$ is the relative impedance of the medium. These decoupled fields satisfy the Helmholtz wave equation separately. The transverse \PW{} solutions $\bm{M}_{\bm{k}}(\bm{r})$ and $\bm{N}_{\bm{k}}(\bm{r})$ of the Helmholtz wave equation (see Eq. (S2)) transform into each other under the action of $\frac{\nabla\times}{k}$, the helicity operator. Thus, we can find solutions $\sqrt{2}\bm{G}_{{\bm{k}},\pm}(\bm{r}) = \bm{M}_{\bm{k}}(\bm{r}) \pm \bm{N}_{\bm{k}}(\bm{r})$ by taking the sum and difference of them. Going back to the electric field, we therefore obtain the solutions
\begin{equation}
	\bm{E}(\bm{r}) = \ee_\pm(\bm{k}_\pm) \e^{\I \bm{k}_\pm\bm{r}}
\end{equation}
for \PW{}s, where $\sqrt{2}\ee_\pm(\bm{k}) = \ee_\text{TE}(\bm{k}) \pm \ee_\text{TM}(\bm{k})$. The factor $\exp(-\I\omega t)$ is dropped throughout this article and the dependence of dispersive quantities on $\omega$ is from now on considered implicitly.

When we assume a primary illumination with a \PW{} of tangential component $\bm{k}_\parallel$ (see \cref{fig:sketch}), we obtain a solution of the following form for each of the half-spaces:
\begin{equation}\label{eq:pwexp}
	\bm{E}(\bm{r}) = \sum_{s=\pm}\sum_{d=\uparrow\downarrow}\sum_{\bm{g}} \alpha_{\bm{g}sd}\ee_s(\bm{k}_{\bm{g}sd})\e^{\I\bm{k}_{\bm{g}sd}\bm{r}}.
\end{equation}
The summation runs over helicities $s$, allowed diffraction orders $\bm{g}$, which are linear combinations of reciprocal lattice vectors, and the parameter $d$, which determines the sign in front of the z-component of the wave vector $\bm{k}_{\bm{g}s,\uparrow\downarrow} = \bm{k}_\parallel \pm \ez\sqrt{k_s^2 - (\bm{k}_\parallel + \bm{g})^2}$. The square root in this expression is taken with positive imaginary part. Only a finite number of diffraction orders are propagating. Thus, the series in \cref{eq:pwexp} can be truncated at a suitable value $g_{\max} > |\bm{g}|$ rendering a finite number of expansion coefficients $\alpha_{\bm{g}sd}$. A desired precision is achieved by taking all the propagating orders and sufficiently many evanescent orders, such that higher evanescent orders decay strongly and have only a negligible contribution to the mutual interaction of adjacent layers. For a linear interaction, the response of the layer can be encoded into four matrices $\Q_{dd'}$ relating the expansion coefficients by
\begin{equation}\label{eq:Qgeneral}
	\begin{pmatrix}
		\alpha_{2,\uparrow} \\
		\alpha_{1,\downarrow} \\
	\end{pmatrix}
	=
	\begin{pmatrix}
		\Q_{\uparrow\uparrow} & \Q_{\uparrow\downarrow} \\
		\Q_{\downarrow\uparrow} & \Q_{\downarrow\downarrow}
	\end{pmatrix}
	\begin{pmatrix}
		\alpha_{1,\uparrow} \\
		\alpha_{2,\downarrow} \\
	\end{pmatrix}
\end{equation}
after arranging the coefficients $\alpha_{\bm{g}sd}$ into vectors $\alpha_{j,d}$ with $j=1, 2$ denoting the half-space below and above the layer. Therefore, the field coefficients on the right side of \cref{eq:Qgeneral} are incoming on the layer and those on the left side are outgoing. The vectors $\alpha_{j,d}$ have the length $2n_g$ with the number of diffraction order $n_g$ and a factor of two for the two helicities. The coupling of two adjacent layers $a$ and $b$ is possible by taking the sum over multiple reflections between these layers, which results in the geometric series
\begin{equation}\label{eq:Qupup}
	\begin{split}
		\Q_{\uparrow\uparrow}
		&= \sum_{n=0}^\infty
		\Q_{\uparrow\uparrow}^b (\Q_{\uparrow\downarrow}^a \Q_{\downarrow\uparrow}^b)^n \Q_{\uparrow\uparrow}^a \\
		&=
		\Q_{\uparrow\uparrow}^b (\mathbbm{1} - \Q_{\uparrow\downarrow}^a \Q_{\downarrow\uparrow}^b)^{-1} \Q_{\uparrow\uparrow}^a\,.
	\end{split}
\end{equation}
Similarly, we obtain the expressions
\begin{equation}\label{eq:Qrest}
	\begin{split}
		\Q_{\uparrow\downarrow}
		= \Q_{\uparrow\downarrow}^b + 
		\Q_{\uparrow\uparrow}^b \Q_{\uparrow\downarrow}^a (\mathbbm{1} - \Q_{\downarrow\uparrow}^b \Q_{\uparrow\downarrow}^a)^{-1} \Q_{\downarrow\downarrow}^b\,, \\
		\Q_{\downarrow\uparrow}
		= \Q_{\downarrow\uparrow}^a + 
		\Q_{\downarrow\downarrow}^a \Q_{\downarrow\uparrow}^b (\mathbbm{1} - \Q_{\uparrow\downarrow}^a \Q_{\downarrow\uparrow}^b)^{-1} \Q_{\uparrow\uparrow}^a\,, \\
		\Q_{\downarrow\downarrow}
		= 
		\Q_{\downarrow\downarrow}^a (\mathbbm{1} - \Q_{\downarrow\uparrow}^b \Q_{\uparrow\downarrow}^a)^{-1} \Q_{\downarrow\downarrow}^b\,.
	\end{split}
\end{equation}
for the remaining three coupled Q-matrices. This new set of Q-matrices describes the joint response of layers $a$ and $b$. In this way, multiple layers can be stacked above each other by iteratively combining adjacent layers \cite{stefanou1998,pendry1974}. For structures having additionally a periodicity in the z-direction, it is possible to use a layer doubling technique to obtain an exponential growth in thickness. This is a convenient way of obtaining the bulk response of thick crystalline structures. This can be achieved when, due to the periodicity, the sets $\Q^a_{dd'}$ and $\Q^b_{dd'}$ are the same. Iteratively plugging in the result of \cref{eq:Qupup,eq:Qrest} back on their right hand side leads to doubling of the number of layers with every iteration.

\subsection{Propagation in chiral media and flat interfaces}
For the expressions in \cref{eq:Qupup,eq:Qrest} to hold, the expansions above layer $a$ and below layer $b$, according to \cref{eq:pwexp}, must be performed with respect to the same origin. To achieve a translation of the origin in a homogeneous medium by a vector $\bm{d}$ we insert an additional layer with Q-matrix entries
\begin{equation}
	\Q_{dd',\bm{g}s\bm{G'}s'} = \delta_{dd'}\delta_{\bm{g}\bm{G'}}\delta_{ss'}\e^{\I\bm{k}_{\bm{g}sd}\bm{d}}\,,
\end{equation}
which apply the right phase factor to each diffraction order for the propagation along $\bm{d}$. Additionally, this vector can also include a translation in x- or y-direction resulting in a lateral shift of the different layers' lattices with respect to each other. Due to the factor $\delta_{dd'}$, the shift of an origin reduces \cref{eq:Qupup,eq:Qrest} to simpler expressions where no matrix inversions are necessary.

Obtaining the Q-matrices of an array of scatterers or an interface is more involved. In the case of an infinitely extended interface, the translational invariance requires the preservation of the tangential component of the \PW{}. Thus, different diffraction orders do not mix. The requirement of continuity for the tangential electric and magnetic field components at the interface leads to the conditions
\begin{subequations}
	\label{eq:interface}
	\begin{eqnarray}
		\bm{E}(x,y,0^-) \times \ez = \bm{E}(x,y,0^+) \times \ez\,, \\
		\bm{H}(x,y,0^-) \times \ez = \bm{H}(x,y,0^+) \times \ez\,.
	\end{eqnarray}
\end{subequations} 
While for non-chiral media it is preferential to solve the interface conditions for transverse electric (TE) and magnetic polarizations (TM), in the chiral case we stick to the helicity basis, since these are the eigenmodes of propagation in chiral media \cite{birula1996}. Still, projections onto different modes are obtained by multiplying with $\ee_{\text{TE}}^\dagger\e^{-\I\bm{k}_\parallel\bm{r}}$ or $\ee_{\text{TE}}^\dagger\times\ez\e^{-\I\bm{k}_\parallel\bm{r}}$ and integrating across the entire x-y-plane, which results in the system of equations
\begin{equation}\label{eq:interface:linsys}
	\begin{pmatrix}
		1 & 1 & -1 & -1 \\
		\frac{k_{z,+,\text{t}}}{k_{+,\text{t}}} &
		-\frac{k_{z,-,\text{t}}}{k_{-,\text{t}}} &
		\frac{k_{z,+,\text{i}}}{k_{+,\text{i}}} &
		-\frac{k_{z,-,\text{i}}}{k_{-,\text{i}}} \\
		Z_\text{t}^{-1} & -Z_\text{t}^{-1} & -Z_\text{i}^{-1} & Z_\text{i}^{-1} \\
		\frac{k_{z,+,\text{t}}}{Z_\text{t}k_{+,\text{t}}} &
		\frac{k_{z,-,\text{t}}}{Z_\text{t}k_{-,\text{t}}} &
		\frac{k_{z,+,\text{i}}}{Z_\text{i}k_{+,\text{i}}} &
		\frac{k_{z,-,\text{i}}}{Z_\text{i}k_{-,\text{i}}}
	\end{pmatrix}
	\begin{pmatrix}
		t_{+s} \\ t_{-s} \\ r_{+s} \\ r_{-s} \\
	\end{pmatrix}
	=
	\begin{pmatrix}
		1 \\
		s\frac{k_{z,s,\text{i}}}{k_{s,\text{t}}} \\
		sZ_\text{i}^{-1} \\
		\frac{k_{z,s,\text{i}}}{Z_\text{i}k_{s,\text{t}}}
	\end{pmatrix}\,,
\end{equation}
where $t_{\pm s} = t_{\pm s}(\bm{k})$ and $r_{\pm s} = r_{\pm s}(\bm{k})$ are the complex transmission and reflection coefficients into helicities $\pm$ and for an incident helicity $s$. The coefficients are functions of the wave vector $\bm{k}$. The relative impedances of the incident and transmitted waves' media are $Z_\text{i}$ and $Z_\text{t}$, respectively. We give full expressions for the transmission and reflection coefficients solving this equation in the supplemental material. These coefficients can be used to build the corresponding Q-matrices as:
\begin{subequations}
	\begin{align}
			\Q_{dd,\bm{g}s\bm{G'}s'} &= \delta_{\bm{GG}'}t_{ss'}(\bm{k}_\parallel + \bm{g}) \\
			\Q_{dd',\bm{g}s\bm{G'}s'} &=  \delta_{\bm{GG}'}r_{ss'}(\bm{k}_\parallel + \bm{g})\quad,~d\neq d'\,.
	\end{align}
\end{subequations}

\subsection{Multi-scattering in bi-periodic arrays}
For the Q-matrices of an array of identical scatterers, the multiple scattering process within it has to be solved. This can be done efficiently by using the T-matrix formalism \cite{waterman1965,xu2013}. Within this formalism, the electric field outside a single isolated object is expanded into \VSW{}s
\begin{equation}\label{eq:vswexp}
	\begin{split}
		\bm{E}(\bm{r}) = \sum_{s=\pm} \sum_{l=1}^\infty \sum_{m=-l}^{l}\left( a_{s,lm} \bm{A}_{s,lm}^{(1)}(k_s\bm{r}) \right.& \\
		\left. + p_{s,lm} \bm{A}_{s,lm}^{(3)}(k_s\bm{r}) \right)&\,,
	\end{split}
\end{equation}
where a basis with multipolar functions of well defined helicity $s$ is chosen instead of the more commonly used electric and magnetic multipoles. Similar to the case of \PW{}s, these basis functions can be constructed by
\begin{equation}
	\bm{A}_{\pm,lm}^{(n)}(k_\pm\bm{r}) = \frac{\bm{N}_{lm}^{(n)}(k_\pm\bm{r}) \pm \bm{M}_{lm}^{(n)}(k_\pm\bm{r})}{\sqrt{2}}
\end{equation}
from TE and TM modes (see Eq. (S3)). Functions of well defined helicity are solutions of the Maxwell equations in chiral media, whereas electric and magnetic modes themselves are not. For the radial dependence in $\bm{A}_{s,lm}^{(1)}$ ($\bm{A}_{s,lm}^{(3)}$) spherical Bessel (Hankel) functions of the first kind are used. The behavior of these functions separates the expansion into incident and scattered fields, respectively. Similarly to the Q-matrices that relate incoming and outgoing waves, the relation of the incident field coefficients $a_{s,lm}$ and scattered field coefficients $p_{s,lm}$ is expressed with
\begin{equation}
	p = \T a\,,
\end{equation}
where $\T$ is a finite matrix after a truncation of the series in $l$. The vectors $a$ and $p$ hold the corresponding expansion coefficients.
The T-matrix itself can be obtained analytically for spherical objects \cite{mie1908,bohren1998,moroz2005}. This is also possible for spherical objects consisting of chiral materials by using $\bm{A}_{s,lm}^{(1)}$ and $\bm{A}_{s,lm}^{(3)}$  \cite{shang2016}. There are various methods to obtain the T-matrix of an arbitrary object, e.g. the extended boundary condition method \cite{waterman1965}, discrete dipole moment method \cite{mackowski2002}, and the use of full-wave solvers \cite{fruhnert2017,demesy2018}. The T-matrix of molecules can be obtained using the output of quantum-mechanical simulations \cite{fernandez-corbaton2020}.

The translation addition theorem for the \VSW{}s \cite{cruzan1962,stein1961,tsang1985} simplifies the treatment of multiple scattering problems significantly. We can describe multiple scattering within the T-matrix formalism by the multiplication with the matrix $\bm{C}^{(3)}(-\bm{R})$ (see supplemental material), resulting in the case of scatterers on a lattice in
\begin{equation}\label{eq:tmatlatstart}
	p_0 = \T \left(a_0 + \sum_{\bm{R}\neq 0} \Cmat^{(3)}(-\bm{R}) p_{\bm{R}}\right)\,.
\end{equation}
In \cref{eq:tmatlatstart} we set the origin of the coordinate system at the position of one scatterer of the array -- since they are all equivalent we can choose one freely -- and use $a_0$ to describe the primary field incident on it, and $p_0$ for the total scattered field emanating from that scatterer.
The sum over all lattice points $\bm{R}$ except the origin includes the re-scattering of the scattered field $p_{\bm{R}}$. Since the illumination is a \PW{} and all scatterers are identical, these fields can be expressed with the coefficients of the scattered field at the origin by $p_{\bm{R}} = \exp(\I \bm{k}_\parallel \bm{R}) p_0$ leading to
\begin{equation}\label{eq:tmatlat}
	p_0 = \left(\mathbbm{1}-\T\sum_{\bm{R}\neq 0} \Cmat^{(3)}(-\bm{R}) \e^{\I\bm{k}_\parallel \bm{R}} \right)^{-1}\T a_0\,.
\end{equation}
The infinite sum over lattice points is numerically challenging, since its direct summation converges slowly. However, inspection of the translation coefficients reveals that the sum has effectively the form
\begin{equation}\label{eq:latsum}
	\sum_{\bm{R}\neq 0} h_l^{(1)}\left(k|\bm{R}|\right) Y_{lm}\left( - \frac{\bm{R}}{|\bm{R}|}\right) \e^{\I \bm{k}_\parallel \bm{R}},
\end{equation}
which is equivalent to the summation needed for the structure constants in low energy electron diffraction theory \cite{pendry1974}. The sum can be converted into two exponentially converging sums by the use of Ewald's method. In the case of a two-dimensional lattice, this method was derived by Kambe \cite{kambe1967}. The scattered field coefficients $p_0$ can be obtained straightforwardly after applying Ewald's method to the infinite summation in \cref{eq:tmatlat}.

The link between the Q-matrix description of the layer and the description in terms of \VSW{}s in \cref{eq:tmatlat} can be established by, first, expanding the incident \PW{}s of \cref{eq:pwexp} in \VSW{}s and, second, summing up the scattered \VSW{}s. The latter results in scattered \PW{}s corresponding to the expansion in \cref{eq:pwexp} due to the lattice symmetry.

The expansion of a \PW{} of well defined helicity in \VSW{}s is a straightforward calculation in the case of normal incidence, i.e. $\bm{k}_\parallel = 0$ \cite[chapter 10.3]{jackson1998}. The result can be generalized to an arbitrary direction of incidence by using the rotation properties of the \PW{}s and \VSW{}s \cite{varshalovich1988,molina-terriza2008} to obtain
\begin{equation}\label{eq:pwtovsw}
	\begin{split}
		&\ee_\pm(\bm{k})\exp(\I \bm{k}\bm{r}) \\&= -\sum_{l=1}^{\infty}
		\sum_{m=-l}^l
		\I^l \sqrt{4\pi(2l+1)} D^l_{m,\pm 1}(0,\theta_{\bm{k}},\varphi_{\bm{k}}) \bm{A}_{\pm,lm}^{(1)}(k\bm{r})\,,
	\end{split}
\end{equation}
and the coefficients for $a_0$ can be directly read off. The functions $D^l_{m,s}$ are components of the Wigner D-matrix $\bm{\mathsf{D}}^l$~\cite{varshalovich1988}.

Finally, the scattered wave summation over the full lattice
\begin{equation}\label{eq:vswtopw}
	\bm{E}_\text{sca}(\bm{r}) = \sum_{\bm{R}} \sum_{l = 1}^{\infty} \sum_{m = -l}^l \sum_{s = \pm} p_{s,lm} A_{s,lm}^{(3)}(k_s (\bm{r}-\bm{R})) \e^{\I \bm{k}_\parallel \bm{R}}
\end{equation}
is possible by using the integral representation of the \VSW{}s derived by Wittmann \cite{wittmann1988} and the direct application of Poisson's formula for the lattice summation \cite{popov2014}. The resulting scattered field is
\begin{equation}
	\bm{E}_{\text{sca},d}(\bm{r}) = \sum_{\bm{g}}
	\sum_{s=\pm}
	E_\text{sca}^{\bm{k}_\parallel+\bm{g},sd} 
	\ee_s(\bm{k}_{\bm{k}_\parallel +\bm{g},s,d})\e^{\I \bm{k}_{\bm{k}_\parallel +\bm{g},s,d}\bm{r}}
\end{equation}
with \PW{} expansion coefficients
\begin{equation}\label{eq:Escacoeff}
	\begin{split}
		E_\text{sca}^{\bm{k}_\parallel+\bm{g},sd} = \sum_{l=1}^\infty\sum_{m=-l}^l p_{s,lm} \frac{2\pi\gamma_{lm}}{A k_s\Gamma_{\bm{k}_\parallel +\bm{g},s}\I^{l+1}} \\
		\left(\frac{m}{\sin\theta}P_l^m(\cos\theta) + s \diffp{}{\theta}P_l^m(\cos\theta)\right)\,,
	\end{split}
\end{equation}

where $A$ is the area of the unit cell, $\gamma_{lm}$ is the normalization factor of the \VSW{}s (see Eq. (S3d)), and $\Gamma_{\bm{k}_\parallel +\bm{g},s} = \sqrt{k_s^2 - (\bm{k}_\parallel + \bm{g})^2}$, with the convention of positive imaginary part for the square root. Furthermore, $\theta = \arccos\frac{\Gamma_{\bm{k}_\parallel + \bm{g},s}}{k_s}$ is the, possibly complex, angle of the \PW{} with the z-direction. We observe  that the \VSW{}s of a given helicity only contribute to \PW{}s of the same helicity and (b) that the \PW{} description emerges naturally after the lattice summation. The scattered field coefficients as defined in \cref{eq:Escacoeff} can be almost directly used as entries for the Q-matrices. The only modification comes from the transition of an incident/scattered to an incoming/outgoing wave description, leading to an addition of 1 to the non-diffracted transmitted \PW{} coefficient.

In summary, all incoming diffraction orders have to be expanded according to \cref{eq:pwtovsw} for the calculation of the full Q-matrix of a single layer of scatterers. Subsequently, the scattered field coefficients can be obtained by \cref{eq:tmatlat}. A major advantage of this calculation method is that the term in brackets in \cref{eq:tmatlat} is invariant under addition of a reciprocal lattice vector. Thus, it only needs to be computed once for each layer of scatterers. Therefore, considering many identical layers is rather simple. Furthermore, the summation with Kambe's method is independent of the scatterers themselves. Finally, the Q-matrix entries can be obtained with \cref{eq:Escacoeff} under consideration of the above mentioned addition of 1.

\subsection{Summary}
The combination of \PW{} and \VSW{} expansions in modes of well defined helicity, layer methods and lattice summation techniques provide a computational efficient way of solving the scattering problem for a diverse number of systems. Typically, the speed enhancement is a factor of 100 to over 1000. Additionally, the method separates the single object properties, as described by its T-matrix, from the lattice influence, which allows efficient parametric studies varying, e.g. the lattice size or geometry, and also offers physical insights. Also, the use of modes with well defined helicity greatly facilitates the use of chiral media in the calculation, and separates these modes in most calculation steps, e.g. in \cref{eq:Escacoeff} and in the translation matrix $\Cmat^{(3)}(-\bm{R})$. It is straightforward to calculate various quantities like transmittance, reflectance, or helicity density analytically from the expansion in \cref{eq:pwexp}. Additionally, the resulting Q-matrices can be used to calculate the band structure \cite{stefanou1998,pendry1974}.

We implemented the algorithm above in Matlab \cite{matlab}. The implementation showed precise agreement with MULTEM \cite{stefanou1998}, when using spheres as scatterers. Since in the case of spheres all formulas are available analytically, we reach a relative difference of the scattered electric field of roughly $10^{-8}$, which is the floating point accuracy. Furthermore, we tested the results for the scattering at arrays of cylinders by comparing with the results obtained with JCMsuite \cite{burger2008}. Here, the the scattered fields were accurate up to relative differences of typically $10^{-4}$ and at maximum $10^{-3}$. We have already used our code for the investigation of helicity preserving cavities featuring arrays of silicon disks \cite{feis2020} and of molecular arrays \cite{fernandez-corbaton2020}. This latter application shows the versatility of the method, which can be used as long as a T-matrix of individual scatterers is available. Reference \cite{fernandez-corbaton2020} shows how to obtain T-matrices of molecular ensembles from quantum-chemical simulations of single molecules.

\section{Exemplary Applications}\label{sec:applic}

We use the algorithm in two examples. First, we exploit mainly the speed up of this calculation method to investigate anti-reflective properties of solar cell nano-coatings made of achiral materials. Second, we investigate the use of dielectric spheres for enhancing circular dichroism signals from a chiral solution. The use of helicity modes is crucial in this second example.

\subsection{Investigation of anti-reflecting coatings for solar cells}
We now exemplify the use of the algorithm described in the last section in a study of an array of dielectric disks that serves as an anti-reflection coating for a solar cell. This example allows us to demonstrate the numerical speed up factor of approximately 500 with respect to finite element method calculations.

\begin{figure*}
	\centering
	\includegraphics[width=\textwidth]{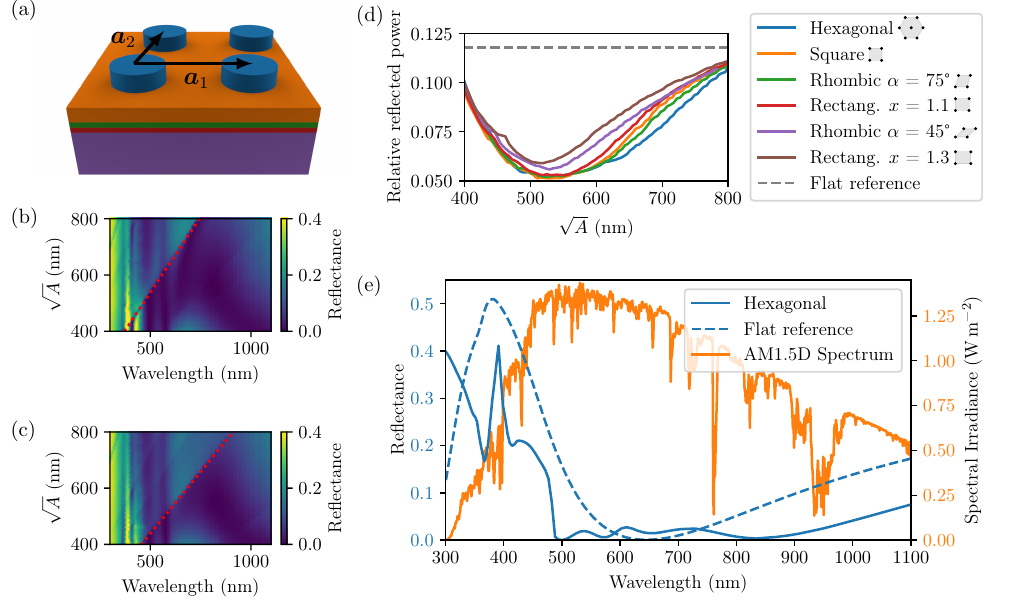}
	\caption{Reflectance from nano-coated solar cells with different Bravais lattices. Panel (a) shows the layout of the solar cell and the lattice vectors of the array. From top to bottom the layers are: ITO (orange), amorphous silicon p-doped (green) and intrinsic (red), and crystalline silicon (purple). Panels (b) and (c) show the reflectance over the square root of the unit cell area $A = |\bm{a}_1 \times \bm{a}_2|$ and the wavelength for a hexagonal and a rectangular lattice with aspect ratio $x = 1.3$, respectively. The red dashed line separates regions with only the zeroth diffraction order propagating in the region below it and with multiple propagating diffraction orders above it. Panel (d) shows the weighted average over the wavelength with the AM1.5D spectrum calculated for six different geometries and compared to a flat reference solar cell. Panel (e) shows the hexagonal lattice with $\sqrt{A} = \SI{520}{\nano\metre}$ in comparison to the flat reference. The AM1.5D spectrum is shown on the same abscissa. In all simulations, the illumination is a plane wave under normal incidence, and we take the average over incoming polarizations.}\label{fig:lat}
\end{figure*}

The layout of the solar cell is shown in \cref{fig:lat}a. It is a heterojunction solar cell and consists of an indium tin oxide (ITO) layer for the top electrode, a p-doped amorphous silicon layer of \SI{4}{\nano\metre}, an intrinsic amorphous silicon layer of \SI{4}{\nano\metre}, and a substrate of crystalline n-doped silicon \cite{leilaeioun2018,schinke2015,pvlighthouse}. In the simulation, the substrate is taken to extend infinitely. To reduce the reflectivity, an array of titanium dioxide (\TIO) \cite{klatio2} cylinders is placed on top of the solar cell. As shown in \cite{fernandez-corbaton2013} for a helicity preserving structure with a rotational axis of degree three or larger, the reflection under normal incidence vanishes completely. The preservation of helicity, i.e. the non-coupling of the two polarization handedness by the structure, can be achieved approximately at least for specific incident wave directions by optimizing the geometrical shape so that the electric and magnetic responses of the structure become equivalent. In \cite{slivina2019} such an optimization has been done for the same -- up to small variations in the permittivities -- solar cell stack that we consider here. The thickness of the ITO layer and the cylinder's height and radius  optimized, resulting in an ITO thickness of \SI{50}{\nano\metre}, a cylinder height of \SI{100}{\nano\metre} and radius of \SI{150}{\nano\metre}. Based on this system, we investigate the influence of the lattice geometry on the anti-reflective properties under normal and oblique incidence.

We compute the T-matrix of the cylinders for the wavelength range between \SI{300}{\nano\metre} and \SI{1100}{\nano\metre} using the commercial finite element solver JCMsuite \cite{pomplun2007, santiago2019}. The maximum expansion order for the T-matrix is set to $l_\text{max} = 8$, which ensures sufficient convergence. The layer setup consists of the cylinder array and four interfaces for the solar cell stack. Q-matrices for the propagation in between these layers are added with the thicknesses mentioned above. Different lattice layouts can be examined efficiently with the algorithm. We choose six layouts, that possess different lattice symmetries: hexagonal, square, rectangular (twice), and rhombic (twice). In the rectangular case, we take side length ratios of $x = 1.1$ and $x = 1.3$, and for the rhombic lattice, we choose angles of $\alpha = \SI{75}{\degree}$ and $\alpha = \SI{45}{\degree}$ between the lattice vectors. We calculate the reflectance for different separations of the cylinders for each of these lattices. This separation is defined by $\sqrt{A}$, where $A$ is the unit cell area. Since the cylinder geometry is not changed, normalizing to $\sqrt{A}$ allows to compare different filling fractions of the surface area of the solar cell. For the calculation of the reflectance, we average over polarizations of the incident light.

Figures \ref{fig:lat}b and \ref{fig:lat}c show the reflectance of solar cells with coatings having a hexagonal lattice or a rectangular lattice with aspect ratio $x = 1.3$ . We observe a clear separation of areas in the range between \SI{400}{\nano\metre} and \SI{900}{\nano\metre}, which is indicated by the red dotted line. Below it, only the zeroth diffraction order propagates, whereas above it the first diffraction order also becomes propagating. The slope of the line is dictated by the lattice geometry. Since the hexagonal lattice has the steepest slope of all lattice geometries, it has the largest area with only the zeroth order propagating. The cylinders are optimized to have a high degree of helicity preservation for this order, thus, it provides low reflectance over a large range of wavelengths. Other lattice geometries, especially if they have a high degree of symmetry, like square arrays, perform comparably. Among the chosen geometries the rectangular lattice with $x=1.3$, whose results are shown in \cref{fig:lat}c, has the earliest onset of multiple diffraction orders. This reduces the range of filling fractions that provide low reflectance. \Cref{fig:lat}d shows the reflectance of all chosen geometries, weighted with the AM1.5D spectrum \cite{gueymard2002} and averaged over all wavelengths. We observe that all geometries have lower reflection than a reference solar cell without coating (gray dashed line). For such flat solar cell the thickness of the ITO layer is optimized to \SI{80}{\nano\metre} \cite{slivina2019} to minimize reflection. We observe in comparison of the different solar cells that the minima with respect to the unit cell size are at approximately the same position and also of similar value. 
Only for lattices with low symmetry this changes slightly. Otherwise, the lattice symmetry mainly influences the width of the minimum in reflected power. Based on the results in \cref{fig:lat}d, the unit cell size is set to \SI{520}{\nano\metre}. \Cref{fig:lat}e shows the reflectance of the solar cell covered with cylinders in a hexagonal array compared to the reflectance of the flat reference solar cell along with the AM1.5D spectrum. The coated solar cell has lower reflectance in a broad range of wavelengths, especially in the range of \SI{400}{\nano\metre} to \SI{580}{\nano\metre} where the  solar irradiance is high. Furthermore, it is considerably lower at long wavelengths. In the range between \SI{580}{\nano\metre} and \SI{740}{\nano\metre}, the reflectance is slightly higher than that of the flat reference.

\begin{figure*}[t]
	\centering
	\includegraphics[width=\textwidth]{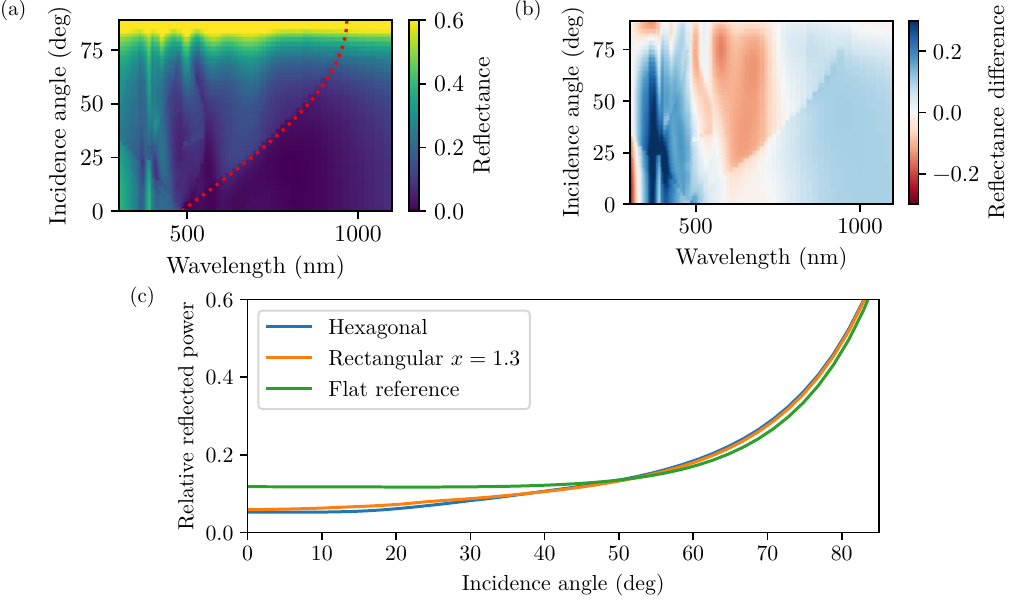}
	\caption{Reflectance of the hexagonal lattice with $\sqrt{A} = \SI{520}{\nano\metre}$ for different angles of incidence. The red dashed line separates regions where only the zeroth order propagates (below) and where multiple propagating diffraction orders propagate (above). Panel (b) shows the comparison to the flat reference and panel (c) the average over wavelength with the AM1.5D spectrum.}\label{fig:angle}
\end{figure*}

To further examine these nano-coatings with low reflectance, we calculate the reflectance under oblique incidence. We take the polarization average as in the case of normal incidence. Additionally, we take the arithmetic average over 24 azimuthal angles of the incident light. \Cref{fig:angle}a shows the reflectance for the hexagonal array. The anti-reflective properties deteriorate slowly for increasing angle of incidence. Similar to the case of different array sizes, the increase in reflectivity is linked with the onset of multiple diffraction orders. This is evident in the area below the red line, where only the zeroth order is propagating. For larger angles, additional diffraction orders change from being evanescent to propagating. The reflectance for oblique incidence in comparison with the flat reference is shown in \cref{fig:angle}b. In regions with blue (red) color the hexagonal lattice has lower (higher) reflection than the flat reference. Upon the onset of multiple diffraction orders, the reflectance of the nano-coated solar cell increases considerably in comparison with the flat solar cell. Finally, we show the averaged reflectance weighted with the AM1.5D spectrum. For angles up to \SI{20}{\degree} the reflectance stays approximately constant for the nano-coated solar cell as well as the flat reference solar cell. With further increasing angle of incidence, the difference between the nano-coated and the flat reference solar cell decreases. At approximately \SI{50}{\degree} the reflectance is equal with and without nano-coating. For large angles of incidence, the flat solar cell is slightly better with respect to its reflectance. The reflectance approaches unity for grazing angles of incidence in all cases.

We finish this example by reporting the computation times for the different cases. All times reported are obtained on a \xeon{} workstation. Unless spheres are used, for which the T-matrix can be calculated analytically, the T-matrix has to be computed numerically, which constitutes an amount of computation time that has to be invested in advance. For the \TIO{} cylinders, this amounts to \SI{4.4}{\hour} for 401 frequencies and a maximum expansion order of $l_\text{max} = 8$. We probed 101 unit cell sizes for each of the six lattice geometries, which took on average \SI{4.6}{\hour}. Thus, a single simulation for one specific array and frequency took \SI{0.4}{\second}. It totals \SI{0.8}{\second} if we add the overhead of the T-matrix computation. The same computation using solely JCMsuite takes approximately \SI{340}{\second} per frequency and lattice geometry, which results in a speed up of approximately 500 in comparison with the \SI{0.8}{\second} for our algorithm and implementation. The computation time of \SI{340}{\second} for a single frequency and lattice would result in a total computation time of approximately 2.6 years for the whole sweep over all lattice types and lattices sizes in the frequency range, which are in total roughly 250.000 different combinations of parameters. The additional simulations of different incident angles also takes approximately the same time per single computation. The polar angle is probed in \SI{2}{\degree} steps and the azimuthal angle in \SI{15}{\degree} steps. The number of azimuthal angles to compute for a full circle coverage depends on the lattice symmetry.

In summary, based on the work of Slivina \textit{et al}, we took cylinders with a high degree of helicity preservation and investigated the influence of the lattice geometry exhaustively. We showed that the main influence for high reflections is by the onset of multiple reflection orders. Thus, for a given filling fraction of the solar cell surface the hexagonal lattice is an appropriate choice, which also shows a robustness against variations of the filling fraction. The nano-coated lattices also show a good performance until quite large angles in comparison to the flat reference solar cell.

\begin{figure}
	\centering
	\includegraphics{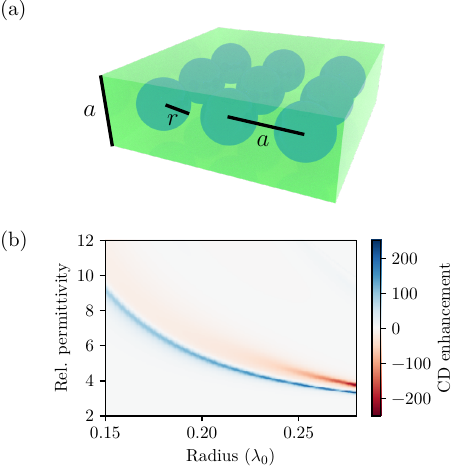}
	\caption{Enhancement of the CD signal from a slab of chiral material with immersed dielectric spheres. Panel (a) shows the structure of the simulated system. The thickness of the slab and the lattice constant of the square sphere array is chosen to be $0.7\lambda_0$ where $\lambda_0$ is the vacuum wavelength. The CD enhancement as a function of the size and permittivity of the spheres is shown in panel (b). The enhancement is taken as the ratio of the CD signal with included spheres to the CD signal of the same volume of material without any inclusions. We observe a resonance of the enhancement with maximum values of 274. However, in close vicinity we find a second resonance with opposite sign with a minimum value of -236.}\label{fig:chiral}
\end{figure}

\subsection{Enhanced circular dichroism signal by dielectric spheres immersed in a chiral solution}

In the previous section we showed the possible speed-up of computations using the algorithm. Now, we present the usage of helicity modes in an environment containing chiral materials. Due to the presence of the chiral material it is highly beneficial to use modes of well-defined helicity, since they are the corresponding eigenmodes. The example aims at the study of the enhancement of the circular dichroism (CD) signal from a slab of chiral material by the inclusion of dielectric spheres. The CD signal is the difference in absorbance of left and right circularly polarized light normalized to.

We model the chiral material by a constant relative permittivity of $\epsilon = 1.333 + 0.001\I$ and a constant chirality parameter of $\kappa = 0.05+0.00015\I$. In this chiral medium, dielectric spheres are arranged on a square lattice with lattice constant $a = 0.7\lambda_0$ as depicted in \cref{fig:chiral}a. Thus, the lattice pitch is just small enough, such that the lattice is in the sub wavelength regime in the chiral material. We vary the radius of the dielectric spheres between $r = 0.15\lambda_0$ and $r = 0.28\lambda_0$. Throughout this example, the incoming plane wave has normal incidence.

We compare the CD signal of this system to a reference system containing the same amount of chiral material but without any spheres. This means that we only consider the CD signal of the bulk material. This differential absorption depends on the amount of material that is considered. Since the enhancement effect is localized in the vicinity of the array, we consider the enhancement within the distance $a$ from the array. The CD enhancement in \cref{fig:chiral}b is given by the ratio of the CD signal of the system with spheres to the reference without spheres.

Within the probed parameter space, we mainly find two sharp resonances where we reach a maximum CD enhancement value of approximately 274. Another resonance where the CD has the opposite sign than the reference slab with a strength of -236 appears close by. The sharp resonances suggest that the enhancement is related to guided modes within the array of spheres. The propagating light effectively encounters a medium with an increased refractive index due to the presence of the spheres. As such, those modes can propagate along the array. The grating structure couples the incident light into this array. Since the system operates close to the sub-wavelength regime, the reflection at the boundaries of the effective waveguide that is formed by the spheres is close to grazing angles. This can lead to a high a degree of helicity preservation which is beneficial for CD enhancement \cite{feis2020}.

\section{Conclusions}\label{sec:concl}
We have presented an algorithm for the rigorous and efficient computation of the electromagnetic response of layered two-dimensional periodic structures. The algorithm is based on the T-matrix formalism, which allows the components of the periodic structures to be particles of arbitrary shape, as well as molecules and molecular ensembles. In addition to an arbitrary number of bi-periodic arrays, any number of additional layers like substrates and superstrates can be included. The use of electromagnetic solutions with well-defined helicity allows the straightforward modeling of isotropic chiral materials in the periodic inclusions, embedding media, and additional layers. We have demonstrated the computational advantages of the presented algorithm with respect to finite-element methods in an exemplary study of nano-coatings for solar cells, showing a speed-up factor of about 500. Additionally, we showed an exemplary simulation of spheres immersed in chiral media where we explicitly made use of the helicity eigenmodes used in our approach.

\section*{Acknowledgments}
This work was supported by the Deutsche Forschungsgemeinschaft (DFG, German Research Foundation) under Germany's Excellence Strategy via the Excellence Cluster 3D Matter Made to Order (EXC-2082/1~--~390761711), via Project RO 3640/12-1 (project number: 413974664), and by the VolkswagenStiftung. D. B. acknowledges support from the Carl Zeiss Foundation via the CZF-Focus@HEiKA program. The authors are grateful to the company JCMwave for their free provision of the FEM Maxwell solver JCMsuite with which the T-matrices of the cylinders discussed in this work were calculated.

\section*{Disclosures}

The authors declare no conflicts of interest.

\bibliography{refs_bibtex}

\end{document}